\documentclass[sigconf,nonacm]{acmart}

\AtBeginDocument{%
  }

\newcommand{\MS}[1]{\textcolor{cyan}{MS: #1}}
\newcommand{\REV}[1]{\textcolor{blue}{#1}}
\newcommand{\JD}[1]{#1}

\setcopyright{rightsretained}
\copyrightyear{2027}
\acmYear{2027}
\acmISBN{}
\acmDOI{}

\acmISBN{}
\acmDOI{}

\begin{document}

\title{A Unified Choice Model of News Selection: Experimental Evidence on Stance, Rank, and Engagement Cues}


\author{Jacopo D'Ignazi}
\email{jacopo.dignazi@upf.edu}
\orcid{0000-0003-2843-5279}
\affiliation{%
  \institution{Universitat Pompeu Fabra}
  \city{Barcelona}
  \country{Spain}
}

\author{Andreas Kaltenbrunner}
\email{andreas.kaltenbrunner@upf.edu}
\orcid{0000-0002-2271-3066}
\affiliation{%
  \institution{Universitat Pompeu Fabra}
  \city{Barcelona}
  \country{Spain}
}

\author{Michele Starnini}
\email{michele.starnini@upf.edu}
\orcid{0000-0002-9161-5339}
\affiliation{%
  \institution{Universitat Pompeu Fabra}
  \city{Barcelona}
  \country{Spain}
}

\author{Vicen\c{c} G\'omez}
\email{vicen.gomez@upf.edu}
\orcid{0000-0001-5146-7645}
\affiliation{%
  \institution{Universitat Pompeu Fabra}
  \city{Barcelona}
  \country{Spain}
}

\renewcommand{\shortauthors}{D'Ignazi et al.}

\begin{abstract}
Social media platforms often display visual engagement cues alongside ranked news articles, yet their influence on clicks remains poorly understood when considered jointly with ranking position and users' political preferences. We propose a unified discrete choice framework using a conditional multinomial logit model to jointly estimate and compare the influence of these factors on news selection. We evaluate the model using 1,480 choices from 370 human participants across four news topics in a controlled study with and without visual engagement cues.
The framework allows us to assess the incremental contribution of each factor through nested-model likelihood-ratio tests.
We find that including visual engagement cues significantly improves model fit across all topics after accounting for rank and stance. 
Engagement cues alone also predict clicks better than either of these features, although the full model achieves the highest held-out log-likelihood. 
Finally, we quantify via counterfactual analysis the impact of ranking position and displayed engagement counts on click probabilities.
These findings can inform platform owners, researchers, and policymakers evaluating how platform design affects the selection of political news.
\end{abstract}




\maketitle

\section{Introduction}
Social media platforms play a central role in political discourse and news consumption. Even small changes in displayed information can alter user behavior~\cite{muchnik2013social}, while ranking algorithms and interface design shape news consumption~\cite{bakshy2015exposure,guess2023social} and have been linked to the amplification of extreme political content, polarization, and misinformation~\cite{dignazi2026rewarding,germano2026ranking,huszar2022algorithmic,avram2020exposure}.


Experiments have shown how ranking position~\cite{ursu2018power,ghose2014examining}, political preferences, and social cues~\cite{engelmann2017comment,dvirgvirsman2019like,epstein2022many} influence user choices. In particular, displaying likes and shares can outweigh partisan preferences, reducing selective exposure to chance levels~\cite{messing2014selective}. However, the joint influence of these factors on political news selection remains poorly understood. For example, \citet{messing2014selective} compare endorsements and partisan sources without varying item position, while \citet{epstein2022many} examine reactions to individual items rather than choices among ranked alternatives. Closer to our approach, \citet{hogg2015disentangling} model position, content, and social signals together, and \citet{krumme2012quantifying} study position and popularity, but both focus on non-political content (science stories and music).

We propose a conditional multinomial logit model that jointly represents political stance, ranking position, and engagement cues as components of utility in a discrete choice setting. Such models have been applied to brand selection~\cite{guadagni1983logit}, product rankings~\cite{ghose2014examining}, network formation~\cite{overgoor2019choosing}, and cultural markets~\cite{krumme2012quantifying}, with extensions for interpretable feature effects~\cite{tomlinson2021learning}.

We illustrate the benefits of the proposed framework through an interventional study examining how visual engagement cues influence news selection.
We use likelihood-ratio tests to assess the contribution of each feature to model fit and cross-validation to compare their predictive power.
 Our work builds on the experimental setup of \citet{dignazi2026rewarding}. We reformulate their clicking model as a special case of our conditional multinomial logit model and analyze data collected with and without visible engagement cues.

Our main contributions are (C1) a unified discrete choice model incorporating political stance, ranking position, and visual engagement cues; (C2) causal evidence that ranking position and the magnitude of displayed engagement counts affect article selection; and (C3) an evaluation of predictive performance and effect sizes, showing that engagement cues alone predict clicks better than either stance or ranking position. \JD{Code and data to reproduce the experiment and all results are available in an anonymized repository.\footnote{\url{https://anonymous.4open.science/r/news-selection-choice-model-B349/}} }

\section{A Discrete Choice Model for News Selection}
\label{sec:model}

The conditional multinomial logit (CML) model describes
the selection of one alternative from a finite choice set through a utility function of observed features. We use it to model which news article a user clicks in a ranked list. For a user $u$ choosing from a set of news articles $A$,
the CML model specifies
\begin{equation}
P(n\mid u,A;\Theta)
=
\frac{\exp(U(u,n;\Theta))}
{\sum_{\tilde n\in A}\exp(U(u,\tilde n;\Theta))},
\qquad n\in A, \nonumber
\end{equation}
where $U(u,n;\Theta)$ is the systematic utility of news
article $n$ for user $u$, parameterized by $\Theta$.

We specify utility using three groups of features:
user--article ideological stances, ranking position, and
visual engagement cues. Stances of a user $u$ and an article $n$
are denoted by $s_u,s_n\in S=\{-2,-1,0 ,1,2\}$,
ranging from extreme left to extreme right.
Let $r_n\in\{1,\ldots,N\}$ denote the ranking position of article $n$, with $r_n=1$ corresponding to the
highest position. Finally, visual engagement cues
are captured by means of the engagement counts $e_n$ associated with article $n$,
which can be, for example, likes or shares.

The utility is a linear combination of features
and their corresponding parameters:
\begin{equation}
\begin{aligned}
U(u,n;\Theta)
&=\Theta^\top\mathbf{x}_{u,n}\\
&=U_a(s_u,s_n;\boldsymbol{\alpha})
 +U_b(r_n;\beta)
 +\delta_c U_c(e_n;\gamma)\\
&=\alpha_{s_u,s_n}
 +\beta(N-r_n)
 +\gamma\delta_c\ln e_n.
\end{aligned}
\nonumber
\end{equation}
Here, $\mathbf{x}_{u,n}$ contains indicators for
user--article stance combinations,
the rank feature $N-r_n$, and the
visibility-dependent engagement feature
$\delta_c\ln e_n$. 
The parameter vector $\Theta$ collects their
corresponding free coefficients.
The indicator $\delta_c$ is $1$ for the
experiment with visible engagement counts and
$0$ otherwise.

The stance parameters form a
$5\times5$ matrix $\boldsymbol{\alpha}$,
where $\alpha_{s_u,s_n}$ captures the preference
of users with stance $s_u$ for articles with stance~$s_n$. Since adding a constant to an entire row
leaves choice probabilities unchanged, we set
$\alpha_{s_u,0}=0$ for every $s_u\in S$,
leaving 20 free parameters.

The rank feature $N-r_n$ has coefficient $\beta\in\mathbb{R}$. A positive $\beta$ favors articles higher in the ranking.
For otherwise identical articles, an article one position higher is $\exp(\beta)$ times as likely to be clicked.

For engagement, we use the feature $\delta_c\ln e_n$,
with coefficient $\gamma\in\mathbb{R}$.
The logarithm reduces the scale of the typically
heavy-tailed engagement counts. Positive values of
$\gamma$ indicate a preference for articles with
higher displayed counts, while $\delta_c=0$ removes
this effect when counts are hidden.
Note that setting $\gamma=0$ yields a reduced model with only stance and position effects, nested within the full model and including the clicking model of~\citet{dignazi2026rewarding} as a special case.


To estimate the model parameters $\Theta$ from a dataset of $m$ observations
$\mathcal{D}=\{(n_k,\{\mathbf{x}_{u_k,n}\}_{n\in A})\}_{k=1}^m$,
we maximize the log-likelihood
\begin{equation}
\begin{aligned}
\mathcal{L}(\Theta)
=\sum_{k=1}^{M}
\left[
U(u_k,n_k;\Theta)
-\ln\sum_{n\in A}
\exp\bigl(U(u_k,n;\Theta)\bigr)
\right].
\nonumber
\end{aligned}
\end{equation}

Because utility is linear in the parameters, the
log-likelihood is concave.
Therefore, any finite local maximum is global.
We fit a common parameter vector across all topics
and separate parameter vectors for each topic.

\section{Experimental setup}\label{sec:data_collection}
Our experiment investigates whether displaying visual engagement cues in ranked news lists influences the selection of news articles.
We conducted two experiments, one without visible engagement cues (E1, March 2026 with 166 participants) and one displaying likes and shares alongside news articles (E2, September 2026 with 204 participants).
Each participant completed four tasks, one on each of the following topics: climate change, immigration, gender policies, and vaccination.
The experimental platform, task design, and news selection
build on \citet{dignazi2026rewarding}, whose implementation
and data are publicly available~\footnote{\url{https://doi.org/10.5281/zenodo.20475404}}.

\begin{figure}[t]
    \centering
    \includegraphics[width=\columnwidth]{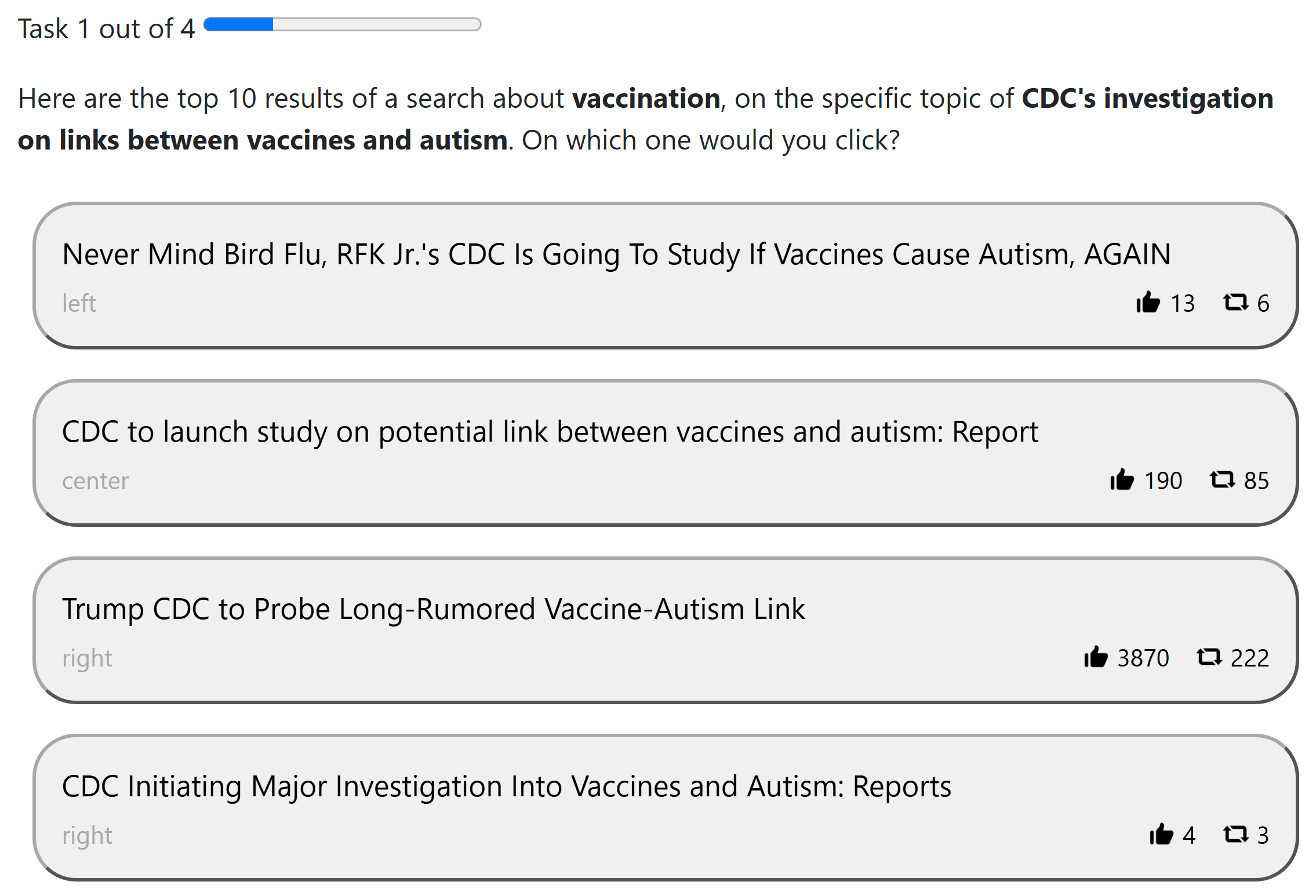}
    \caption{Example of a randomized ranked news list users see during one task of our experiment (E2). Engagement metrics are shown at the bottom right of each box. Users can scroll to see all 10 news items before deciding which one to click.}
    \label{fig:platform_ranking}
\end{figure}

\subsubsection*{User Recruitment}
We recruited participants from the US through Prolific,
requiring an approval rate above $90\%$.
We used quotas to balance self-reported political
affiliation (Republican, Democratic, or Independent)
and gender (male or female). No quotas were imposed on
other demographic characteristics. The age distribution of participants was centered around 40-50 years.

To mitigate the participation of bots, we included an attention check which asked the users to retrieve a number from an external web page\footnote{\url{https://justflipacoin.com}}.
Participants were then asked to write that number on a second page, and those who failed to write one in the expected range were immediately excluded from the experiment.   
Approximately $80\%$ of recruited participants passed this check. Participants who passed and accepted the study conditions completed the four tasks within 40 minutes and received approximately \pounds1.5
(equivalent to \pounds9 per hour based on the median completion time).


\subsubsection*{News Selection}
For each of the four topics, we selected 10 articles through
Ground News\footnote{\url{https://ground.news}},
two from each of the five ideological
categories defined in Section~\ref{sec:model}.
To reduce the variation associated with differences in the events covered, we selected articles from different news outlets reporting on the same event. 


\subsubsection*{Platform Design}
We implemented the experiment using oTree, with an interface resembling a social media news feed. In each task, participants first read a brief description of the topic and reported their stance on a five-point scale, from strongly left to strongly right. They then selected one article from a scrollable ranked list of 10 articles. The order of the four topics varied across participants.

Article headlines and coarse stance labels (left, center, or right) were always visible, while the news source and the distinction between moderate and extreme stances were hidden. In E2, each article also displayed its number of likes and shares (Fig.~\ref{fig:platform_ranking}). These counts were hidden in E1.



For each participant and topic, article positions were
randomized. In E2, pairs of like and share counts were also
randomly assigned to articles, independently of their
ranking positions. These values were derived from the
engagement recorded in~\citep{dignazi2026rewarding} and
rescaled exponentially to make differences between low and high counts more apparent to participants. The smallest values were on the order of some units and the largest of $O(1000)$.

\section{Results}\label{sec:model_evaluation}
We first assess the stability of parameter estimates, followed by likelihood-ratio tests to evaluate the contribution of each model component. We then compare predictive performance through cross-validation and, finally, quantify the effects of ranking position and displayed engagement counts on click probabilities.

\subsubsection*{Stability Test}\label{sec:stability}
To assess the stability of parameter estimates as the dataset size increases, we first fit the model to all E2 data to obtain reference parameters \(\widehat{\Theta}_\text{ref}\).
We consider dataset sizes in the range from 10 to 816 clicks. For each size, we generate 1,000 synthetic datasets and re-estimate the model parameters.
We then compute the relative absolute differences of the estimates from the reference values, and report the mean and standard deviation of such differences. 
For stance, we convert the coefficients into
normalized preference probabilities by applying a softmax
over article stances for each user stance. This accounts
for the fact that stance utilities are defined up
to an additive constant. For rank and engagement, we
compute relative absolute differences directly from
$\beta$ and~$\gamma$.

These estimation errors decrease rapidly as dataset size
increases. For datasets of 200 clicks, the relative absolute differences
are approximately 20--40\% for stance parameters and
10--20\% for rank and engagement parameters.
With the largest dataset, these values decrease to
10--20\% and 5--10\%, respectively.
Across all dataset sizes, $\beta$ shows slightly larger
deviations than $\gamma$. 
The estimates show little further improvement beyond approximately 600 clicks.

\subsubsection*{Ablation Analysis}
\label{sec:ablation_analysis}

\begin{table}[t]
\centering
\caption{Likelihood-ratio tests comparing the full model with the model that omits one component ($p$-values Holm-adjusted across all 15 tests). Omitting any of the components is significant in the pooled data, and engagement is significant on every topic. Results on stance and rank are mixed.}
\label{tab:ablation}
\small
\setlength{\tabcolsep}{3pt}
\begin{tabular}{@{}lccccc@{}}
\toprule
Omitted Component & Overall & Gender & Vaccines & Immigration & Climate \\
\midrule
Stance & $<10^{-5}$ & 0.041 & $<10^{-5}$ & 0.127 & $<10^{-5}$ \\
Rank & $<10^{-5}$ & $\sim 10^{-5}$ & $<10^{-5}$ & $\sim 10^{-5}$ & 0.127 \\
Engagement (Likes) & $<10^{-5}$ & $<10^{-5}$ & $<10^{-5}$ & $\sim 10^{-5}$ & $<10^{-5}$ \\
\bottomrule
\end{tabular}
\end{table}

We assess the contribution of each model component using a
balanced pooled dataset combining E1 and E2 (selecting the first 166 users of E2 in chronological order), for a total of 1,328 clicks with the same news articles in both conditions. For the overall dataset and each topic, we
compare the full model with three nested models, each
omitting one component: stance, rank, or engagement.
We use likelihood-ratio tests, with 20 degrees of freedom
for stance and one for rank and engagement.
Table~\ref{tab:ablation} reports the resulting $p$-values,
adjusted for multiple comparisons using Holm's method.

All three components significantly improve model fit
across the full dataset. Engagement is significant for
every topic, with $\gamma>0$ in all cases, while stance
and rank show more variation across topics.
In particular, stance is not significant for immigration
($p=0.127$), nor is rank for climate change ($p=0.127$).

The engagement component differs from the others because
it enters the utility only when counts are visible
($\delta_c=1$). Thus, E1 contributes to estimating the
common stance and rank parameters, while $\gamma$ is
identified through the displayed counts in E2.

Since ranking positions and engagement counts were
randomized independently, both effects admit a causal
interpretation. Articles appearing higher in the ranking
are more likely to be selected, as are articles displaying
higher engagement counts in E2.

\subsubsection*{Predictivity analysis}

The previous likelihood-ratio tests assess whether
each feature significantly improves model fit.

\begin{figure}[!t]
    \centering
        \centering
        \includegraphics[width=\columnwidth]{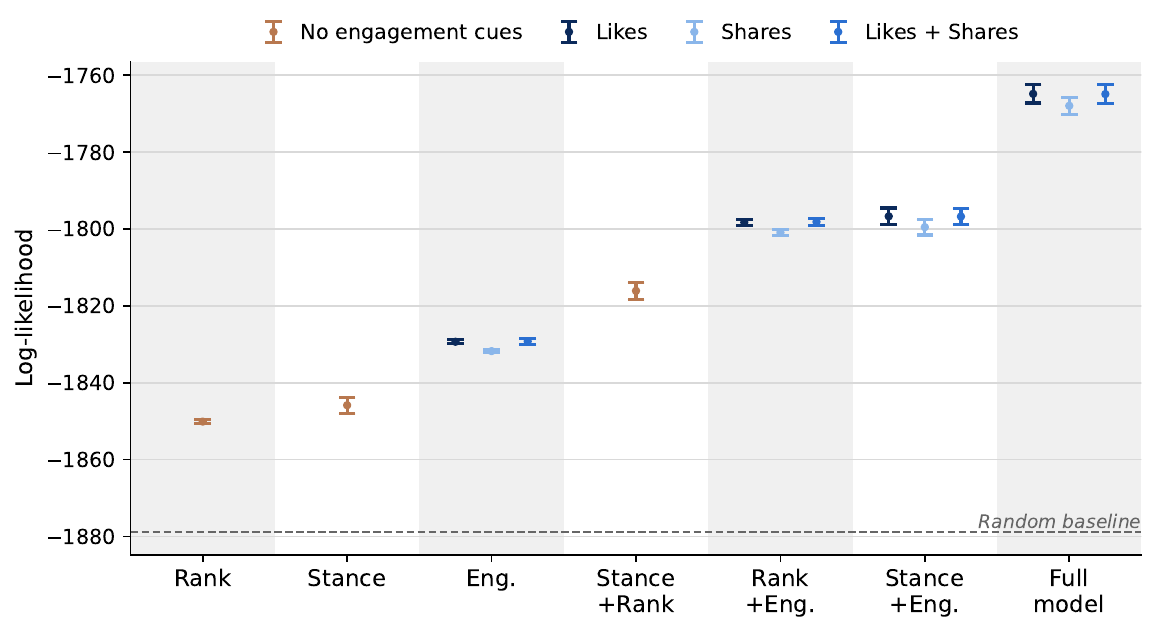}
\caption{Mean and standard deviation of out-of-sample
log-likelihood over 10 repetitions of 10-fold
cross-validation on E2. Colors distinguish the
engagement features included in each model.
The dashed line indicates the uniform random baseline.}
    \label{fig:CV_results}
\end{figure}

We now examine how well different combinations of
features predict unseen clicks using cross-validation.
We evaluate each model specification on E2 data using
10 repetitions of 10-fold cross-validation. For each
repetition, we sum the out-of-sample log-likelihood
across the 10 folds. Figure~\ref{fig:CV_results}
reports the mean and standard deviation across repetitions.

Within each repetition, all four choices from the same
participant were assigned to the same fold, ensuring
that participants in the test set were not included
in the training set.

As a baseline, we consider uniform random selection,
a model that assigns probability $1/10$ to each article.
For the 816 clicks in E2, this corresponds to a log-likelihood of $-816\ln(10)\approx -1879$,
shown as a dashed line in Fig.~\ref{fig:CV_results}.

For the engagement component, we consider three
specifications: Likes, using the number of likes;
Shares, using the number of shares; and Likes + Shares,
using the sum of both counts. Each specification
uses a single engagement feature and coefficient.

Engagement cues alone achieve higher predictive
performance than either stance or rank alone.
The differences between engagement measures are small, with likes yielding higher log-likelihoods than shares.
Using their sum offers no clear improvement over likes alone. The highest out-of-sample log-likelihood is obtained when stance, rank, and engagement are included together.

\subsubsection*{Counterfactual Click Probabilities}


To quantify the effects of ranking position and displayed engagement on click probabilities, we use the model fitted on E2 to compute counterfactual predictions for the 816 rankings presented to participants. For each ranking, we first swap the positions of the top- and bottom-ranked articles. We then consider a separate counterfactual in which the articles with the lowest and highest engagement exchange their like and share counts. In each case, we compare the predicted click probability of the article initially at the bottom or with the lowest engagement, before and after the swap, keeping all other features unchanged. We report the interquartile ranges (25th--75th percentiles)
of these probabilities and their relative changes across rankings.


Moving an article from the bottom to the top increases its predicted click probability from 3.5--7.7\% to 9.2--18.9\%, corresponding to relative increases of 141--162\%. Similarly, exchanging the lowest and highest engagement counts increases the predicted probability from 3.9--8.4\% to 12.5--24.4\%, with relative increases of 158--268\%.


The observed clicks show a similar pattern. Bottom-ranked articles received 7.8\% of clicks, compared with 17.8\% for top-ranked articles. Articles displaying the lowest like counts received 7.2\% of clicks, compared with 21\% for those displaying the highest counts.


\section{Discussion}
We have presented a conditional multinomial logit model that jointly accounts for political stance, ranking position, and visual engagement cues in news selection. The model includes the clicking model of \citet{dignazi2026rewarding} as a special case and incorporates engagement cues into the same likelihood-based framework.

Cross-validation shows that engagement cues alone predict clicks better than stance or rank, while combining all components yields the highest out-of-sample log-likelihood. 

The fitted model estimates how click probabilities change when an article's ranking position or displayed engagement counts are modified. Since both features were randomized in our experiment, these counterfactual estimates can be interpreted causally. 
These predictions show that moving an article from the bottom to the top of a ranking more than doubles its click probability, while assigning it the highest instead of the lowest engagement counts roughly triples it.

Overall, our findings highlight the importance of visual engagement cues in shaping political news selection, beyond ranking position and ideological preferences.
Our results suggest that the way platforms display popularity information can affect which political news articles attract attention. Whether these effects persist in more realistic browsing environments or have implications for news diversity and political polarization remains an open question.

\paragraph{Limitations.}
The two experimental conditions were conducted six
months apart (March and September 2026), without random
assignment of participants between conditions.
Both conditions used the same articles and recruitment criteria, but differences between the two samples may remain.
Our causal evidence for engagement comes from the randomized counts within E2 and concerns their displayed magnitude, not the effect of making engagement cues visible.


\paragraph{Ethical Use of Data and Informed Consent.}
The study followed ACM's Publications Policy on Research
Involving Human Participants and Subjects (2021) and
applicable institutional and disciplinary ethical standards.
All procedures received ethics board approval, and
participants provided informed consent. The consent form
stated that participation ``does not entail risks greater
than those ordinarily encountered in daily life''.
Political affiliations previously disclosed to Prolific
were used for quota sampling.

\bibliographystyle{ACM-Reference-Format}
\bibliography{references}

@article{muchnik2013social,
  author  = {Muchnik, Lev and Aral, Sinan and Taylor, Sean J.},
  title   = {Social Influence Bias: A Randomized Experiment},
  journal = {Science},
  volume  = {341},
  number  = {6146},
  pages   = {647--651},
  year    = {2013},
  doi     = {10.1126/science.1240466}
}

@article{bakshy2015exposure,
  author  = {Bakshy, Eytan and Messing, Solomon and Adamic, Lada A.},
  title   = {Exposure to Ideologically Diverse News and Opinion on {Facebook}},
  journal = {Science},
  volume  = {348},
  number  = {6239},
  Xpages   = {1130--1132},
  year    = {2015},
  doi     = {10.1126/science.aaa1160}
}

@article{guess2023social,
  author  = {Guess, Andrew M. and Malhotra, Neil and Pan, Jennifer and Barber{\'a}, Pablo and Allcott, Hunt and others},
  title   = {How Do Social Media Feed Algorithms Affect Attitudes and Behavior in an Election Campaign?},
  journal = {Science},
  volume  = {381},
  number  = {6656},
  pages   = {398--404},
  year    = {2023},
  doi     = {10.1126/science.abp9364}
}

@inproceedings{dignazi2026rewarding,
  author    = {D'Ignazi, Jacopo and Fraxanet, Emma and Kaltenbrunner, Andreas and Le Mens, Ga{\"e}l and Germano, Fabrizio and G{\'o}mez, Vicen{\c{c}}},
  title     = {Rewarding Engagement and Personalization in Popularity-Based Rankings Amplifies Extremism and Polarization},
  xbooktitle = {Proceedings of the 32nd ACM SIGKDD Conference on Knowledge Discovery and Data Mining V.2 (KDD)},
  booktitle = {Proceedings of the 32nd ACM SIGKDD Conference},
  pages     = {875--886},
  year      = {2026},
  doi       = {10.1145/3770855.3818037}
}

@article{huszar2022algorithmic,
  author  = {Husz{\'a}r, Ferenc and Ktena, Sofia Ira and O'Brien, Conor and Belli, Luca and Schlaikjer, Andrew and Hardt, Moritz},
  title   = {Algorithmic Amplification of Politics on {Twitter}},
  journal = {Proceedings of the National Academy of Sciences},
  volume  = {119},
  number  = {1},
  pages   = {e2025334119},
  year    = {2022},
  doi     = {10.1073/pnas.2025334119}
}

@article{avram2020exposure,
  author  = {Avram, Mihai and Micallef, Nicholas and Patil, Sameer and Menczer, Filippo},
  title   = {Exposure to Social Engagement Metrics Increases Vulnerability to Misinformation},
  journal = {Harvard Kennedy School Misinformation Review},
  volume  = {1},
  number  = {5},
  year    = {2020},
  doi     = {10.37016/mr-2020-033}
}

@article{ursu2018power,
  author  = {Ursu, Raluca M.},
  title   = {The Power of Rankings: Quantifying the Effect of Rankings on Online Consumer Search and Purchase Decisions},
  journal = {Marketing Science},
  volume  = {37},
  number  = {4},
  pages   = {530--552},
  year    = {2018},
  doi     = {10.1287/mksc.2017.1072}
}

@article{ghose2014examining,
  author  = {Ghose, Anindya and Ipeirotis, Panagiotis G. and Li, Beibei},
  title   = {Examining the Impact of Ranking on Consumer Behavior and Search Engine Revenue},
  journal = {Management Science},
  volume  = {60},
  number  = {7},
  pages   = {1632--1654},
  year    = {2014},
  doi     = {10.1287/mnsc.2013.1828}
}

@article{engelmann2017comment,
  author  = {Engelmann, Ines and Wendelin, Manuel},
  title   = {Comment Counts or News Factors or Both? {I}nfluences on News Website Users' News Selection},
  journal = {International Journal of Communication},
  volume  = {11},
  pages   = {2501--2519},
  year    = {2017}
}

@article{dvirgvirsman2019like,
  author  = {Dvir-Gvirsman, Shira},
  title   = {I Like What {I} See: Studying the Influence of Popularity Cues on Attention Allocation and News Selection},
  journal = {Information, Communication \& Society},
  volume  = {22},
  number  = {2},
  pages   = {286--305},
  year    = {2019},
  doi     = {10.1080/1369118X.2017.1379550}
}

@misc{epstein2022many,
  author        = {Epstein, Ziv and Lin, Hause and Pennycook, Gordon and Rand, David},
  title         = {How Many Others Have Shared This? {E}xperimentally Investigating the Effects of Social Cues on Engagement, Misinformation, and Unpredictability on Social Media},
  year          = {2022},
  eprint        = {2207.07562},
  archivePrefix = {arXiv},
  primaryClass  = {cs.SI},
  doi           = {10.48550/arXiv.2207.07562}
}

@article{messing2014selective,
  author  = {Messing, Solomon and Westwood, Sean J.},
  title   = {Selective Exposure in the Age of Social Media: Endorsements Trump Partisan Source Affiliation When Selecting News Online},
  journal = {Communication Research},
  volume  = {41},
  number  = {8},
  pages   = {1042--1063},
  year    = {2014},
  doi     = {10.1177/0093650212466406}
}

@article{hogg2015disentangling,
  author  = {Hogg, Tad and Lerman, Kristina},
  title   = {Disentangling the Effects of Social Signals},
  journal = {Human Computation},
  volume  = {2},
  number  = {2},
  year    = {2015},
  doi     = {10.15346/hc.v2i2.4}
}

@article{krumme2012quantifying,
  author  = {Krumme, Coco and Cebrian, Manuel and Pickard, Galen and Pentland, Sandy},
  title   = {Quantifying Social Influence in an Online Cultural Market},
  journal = {PLoS ONE},
  volume  = {7},
  number  = {5},
  pages   = {e33785},
  year    = {2012},
  doi     = {10.1371/journal.pone.0033785}
}

@article{guadagni1983logit,
  author  = {Guadagni, Peter M. and Little, John D. C.},
  title   = {A Logit Model of Brand Choice Calibrated on Scanner Data},
  journal = {Marketing Science},
  volume  = {2},
  number  = {3},
  pages   = {203--238},
  year    = {1983},
  doi     = {10.1287/mksc.2.3.203}
}

@inproceedings{overgoor2019choosing,
  author    = {Overgoor, Jan and Benson, Austin R. and Ugander, Johan},
  title     = {Choosing to Grow a Graph: Modeling Network Formation as Discrete Choice},
  booktitle = {Proceedings of the World Wide Web Conference (WWW)},
  pages     = {1409--1420},
  year      = {2019},
  doi       = {10.1145/3308558.3313662}
}

@inproceedings{tomlinson2021learning,
  author    = {Tomlinson, Kiran and Benson, Austin R.},
  title     = {Learning Interpretable Feature Context Effects in Discrete Choice},
  booktitle = {Proceedings of the 27th ACM SIGKDD Conference on Knowledge Discovery \& Data Mining (KDD)},
  pages     = {1582--1592},
  year      = {2021},
  doi       = {10.1145/3447548.3467250}
}

@article{germano2026ranking,
  title={Ranking for engagement: How social media algorithms fuel misinformation and polarization},
  author={Germano, Fabrizio and G{\'o}mez, Vicen{\c{c}} and Sobbrio, Francesco},
  journal={Journal of Public Economics},
  volume={255},
  pages={105589},
  year={2026},
  publisher={Elsevier}
}

\end{document}